\def\ARXIV{1} 
\def\REVIEW{0} 
\def\BLIND{0} 
\def\BBL{0} 
\def\NOAUTHORS{0} 

\ifnum\BLIND=0
    \documentclass[authoryear,preprint,12pt]{elsarticle}
\else
    \documentclass[authoryear,doubleblind,preprint,12pt]{elsarticle}
\fi

\usepackage{amssymb}

\usepackage{booktabs}
\usepackage[hidelinks,breaklinks]{hyperref}
\usepackage{multirow}
\usepackage{xcolor}
\usepackage{threeparttable} 
\usepackage{siunitx} 
\usepackage[normalem]{ulem} 
\robustify\uline
\usepackage{subcaption}

\usepackage{soul}
\usepackage{amsmath}
\usepackage{longtable}
\usepackage[normalem]{ulem}
\useunder{\uline}{\ul}{}
\usepackage{float}
\usepackage[normalem]{ulem}

\ifnum\REVIEW=1
    \usepackage{lineno}
\fi

\ifnum\ARXIV=0
    \journal{International Journal of Forecasting}
\else
    \journal{arXiv}
\fi

\begin{document}
\begin{frontmatter}

\title{Not feeling the buzz: Correction study of mispricing and inefficiency in online sportsbooks}

\ifnum\NOAUTHORS=1
    \author{}
\else
    \author{Lawrence Clegg}
    \ead{tb21144@bristol.ac.uk}
    \author{John Cartlidge\corref{cor1}}
    \ead{john.cartlidge@bristol.ac.uk}
    \cortext[cor1]{Corresponding author}
    \affiliation{organization={Department of Computer Science, University of Bristol},
            city={\\Merchant Venturers Building, Woodland Road, Bristol},
            postcode={BS8 1UB}, 
            country={UK}}
\fi

\begin{abstract}
We present a replication and correction of a recent article (Ramirez, P., Reade, J.J., Singleton, C., Betting on a buzz: Mispricing and inefficiency in online sportsbooks, International Journal of Forecasting, 39:3, 2023, pp. 1413--1423, doi: 10.1016/\-j.ijforecast.2022.07.011).
RRS measure profile page views on Wikipedia to generate a ``buzz factor'' metric for tennis players and show that it can be used to form a profitable gambling strategy by predicting bookmaker mispricing. 
Here, we use the same dataset as RRS to reproduce their results exactly, thus confirming the robustness of their mispricing claim. 
However, we discover that the published betting results are significantly affected by a single bet (the ``Hercog'' bet), which returns substantial outlier profits based on erroneously long odds. 
When this data quality issue is resolved, the majority of reported profits disappear and only one strategy, which bets on ``competitive'' matches, remains significantly profitable in the original out-of-sample period. While one profitable strategy offers  weaker support than the original study, it still provides an indication that market inefficiencies may exist, as originally claimed by RRS.
As an extension, we continue backtesting after 2020 on a cleaned dataset. Results show that (a) the ``competitive'' strategy generates no further profits, potentially suggesting markets have become more efficient, and (b) model coefficients estimated over this more recent period are no longer reliable predictors of bookmaker mispricing. We present this work as a case study demonstrating the importance of replication studies in sports forecasting, and the necessity to clean data. We open-source release comprehensive datasets and code.
\end{abstract}


\ifnum\ARXIV=1
\else
    \ifnum\BLIND=1
    \else
        \begin{highlights}
        \item Research highlight 1 - we perform a strict replication of the ``WikiBuzz'' bookmaker mispricing model and betting strategy introduced by \cite{BettingOnABuzz}.
        \item Research highlight 2 - we show that the impressive Bet365 betting profits reported by \cite{BettingOnABuzz} largely result from a single, extremely profitable bet that is only possible due to a data error in available odds. 
        \item Research highlight 3 - we provide corrected results and demonstrate that bookmaker mispricing and market inefficiencies remain.
        \item Research highlight 4 - we extend the work by exploring profitable model performance over a more recent period, post 2020. Results suggest that markets have become more efficient and WikiBuzz and RankDist factors are no longer reliable predictors of bookmaker mispricing. 
        \item Research highlight 5 - we open-source release comprehensive datasets and code at \href{https://github.com/Faxulous/notFeelingTheBuzz}{https://github.com/Faxulous/notFeelingTheBuzz}.
        \end{highlights}
    \fi
\fi


\end{frontmatter}

\ifnum\REVIEW=1
    \linenumbers
\fi


\section{Introduction}\label{sec:intro}
\noindent
In a recent {\em International Journal of Forecasting} (IJF) article, Ramirez, Reade, and Singleton (RRS) introduced a novel ``Wikipedia Relative Buzz Factor'' metric \citep{BettingOnABuzz}. This ``WikiBuzz'' metric, which measures excess pre-match page views of tennis players' profiles on Wikipedia, is shown to significantly predict mispricing of bookmakers' odds for tennis match outcomes. It is then used as part of a betting strategy, to test for market inefficiency. The authors showed that ``a strategy of betting on players who received more pre-match buzz than their opponents can generate substantial profits''; and ``returns on investment from applying the model and betting strategy were sustained and substantial, including when using only the odds of Bet365, the world’s highest revenue online sportsbook'' \citep{BettingOnABuzz}. The intriguing claims of ``sustained'' and ``substantial'' betting returns when using only readily available odds from a single online sportsbook, motivated us to perform a replication study to independently verify these findings. Verification is particularly important given that the reported return on investments (ROI), which reach as high as 29.38\%, are much higher than reported elsewhere in the tennis betting literature, where ROIs of 3.56\% \citep{angelini2022weighted}, 3.8\% \citep{knottenbelt2012common}, and 4.35\% \citep{sipko2015machine} have been reported.

In this work, we write and release new Python code, which follows the same algorithm implemented by RRS in Stata code. Using RRS' original dataset, we were able to {\em exactly replicate all results} presented by \cite{BettingOnABuzz}.\footnote{We describe some minor differences in our replication, resulting from e.g., a typo, and handling of floating point precision, but these are all inconsequential.} This outcome independently verifies that the profits reported by \citeauthor{BettingOnABuzz} are generally accurate when backtested on their specific dataset. Furthermore, it also verifies that our algorithmic implementation of the published method is correct. 
By following RRS' application of the \cite{mincer69eval} forecast evaluation framework, we also find that ``WikiBuzz'' and ``RankDist'' coefficients are significantly non-zero. We present this as evidence to confirm that RRS' conclusion of the existence of bookmaker mispricing is robust.

However, for RRS' best-performing betting models, we find that a large proportion of profits are generated by a single bet (the ``Hercog'' bet) on Bet365 sportsbook, which results from an inconsistency (i.e., an {\em error}) in the odds data provided by {\tt tennis-data.co.uk}. Once the dataset is cleaned to remove such inconsistencies, we find that only one of the proposed betting strategies that use odds from Bet365 remains significantly profitable. 

We extend the work of RRS by testing the remaining profitable strategy with an additional 3 years of data (covering the period 2020--2023) and find that no further profits are generated. We also show that new maximum likelihood parameter estimates for the WikiBuzz metric for the new data period are no longer significant, and therefore WikiBuzz and RankDist factors are no longer reliable predictors of bookmaker mispricing.

The rest of this paper is organised as follows. In Section~\ref{sec:replication}, we perform a strict replication using the original dataset to independently verify results. Then, in Section~\ref{sec:corrections}, we demonstrate that the profitable returns for the strategy that uses Bet365 odds are generated by an abnormal windfall from a single bet that is directly caused by an error in odds data. Once we correct for this error, only one strategy using Bet365 odds remains profitable. In Section~\ref{sec:extension}, we explore the performance of the remaining profitable strategy over an extended and cleaned dataset, with results tentatively suggesting that bookmaker mispricing and market inefficiencies have diminished since 2020. In Section~\ref{sec:discussion}, we discuss the need to revisit other sports forecasting works in light of our findings. Finally, Section~\ref{sec:conclusion} concludes with a call-to-arms for replication studies. We make all code and new datasets available open source for others to replicate and reproduce our findings.

\section{Replication}\label{sec:replication}
\noindent
We present a strict replication of RRS, using the same third-party data from {\tt tennis-data.co.uk}.\footnote{
Link to RRS' data and code: \href{https://github.com/philiprami/betting_on_a_buzz}{https://github.com/philiprami/betting\_on\_a\_buzz}.
}
The dataset contains WTA data sourced from {\tt tennis-data.co.uk} for the period July 2015 to February 2020, inclusive; which includes information on every tennis match (competitors, competitor rankings, tournament, location, date, result, etc.) and a selection of pre-match betting odds taken from the major bookmakers, such as Bet365, as well as the market's average odds and best odds taken from the $K$ individual bookmakers listed on {\tt oddsportal.com} (where $K$ is normally in the range 40-60).\footnote{A full description of the tennis data is available online: \href{http://www.tennis-data.co.uk/notes.txt}{tennis-data.co.uk/notes.txt}.} 
The dataset also contains each competitor's Wikipedia page view metrics,\footnote{Column headers: \emph{wiki\_yesterday\_w, wiki\_yesterday\_l, wiki\_med365\_w, wiki\_med365\_l}; for the winner's views yesterday, loser's views yesterday, annual median views for the winner, and annual median views for the loser, respectively.} which were calculated using the Wikimedia Foundation Pageview API.\footnote{API documentation: \href{https://wikitech.wikimedia.org/wiki/Analytics/AQS/Pageviews}{https://wikitech.wikimedia.org/wiki/Analytics/AQS/Pageviews}}

We pre-process the data by removing rows with missing values.
The data is then split into two sets, with years 2016-2018, inclusive, used to fit the model (the in-sample training set); and 2019 to February 2020, inclusive, used to assess betting returns (the out-of-sample test set).

Here, we develop an alternative Python implementation\footnote{
New code and data: \href{https://github.com/Faxulous/notFeelingTheBuzz}{https://github.com/Faxulous/notFeelingTheBuzz}
}, and make use of the {\tt linearmodels} Python library\footnote{linearmodels Python package, \href{https://bashtage.github.io/linearmodels}{https://bashtage.github.io/linearmodels}} to estimate Model (\ref{eq:errormodel}) for mispricing, and both Model (\ref{eq:pm}) and Model (\ref{eq:pmword}) for outcome prediction and inefficiency testing.

\subsection{Bookmaker mispricing}\label{sec:model}
\noindent
RRS present results of bookmaker mispricing in \cite{BettingOnABuzz}, Section 3.1, Tables 1, 2, 3. Here, we briefly outline the model used by RRS. For more details on model rationale, we refer the reader to the original article. 

The authors first apply a \cite{mincer69eval} forecast evaluation framework to estimate the conditional mean effects on the bookmakers' odds implied probability forecast errors:
\begin{equation}\label{eq:errormodel}
    e_{ij} = \alpha + \beta_1z_{ij} + \beta_2\text{RankDist}_{ij} + \beta_3\text{WikiBuzz}_{ij} + \psi_{S(j)} + \phi_{T(j)} + \varepsilon_{ij},
\end{equation}
\noindent
 where the error $e_{ij}$ is the distance between the odds implied probability and $0$ if the outcome was a loss, and $e_{ij}$ is the distance between the odds implied probability and $1$ if the outcome was a win; $z_{ij}$ is the odds implied probability (i.e., the market's estimate of outcome $\widetilde{y}_{ij}$, using average odds of bookmakers); $\textup{RankDist}_{ij}$ is the distance between rankings of opponents (i.e., the relative strength of opponents); $\textup{WikiBuzz}_{ij}$ is \citeauthor{BettingOnABuzz}'s bespoke Wiki Relative Buzz Factor (i.e., the relative pre-match interest surrounding opponents); and $\hat{\alpha}, \hat{\beta_{1}}, \hat{\beta_{2}}, \hat{\beta_{3}}$ are parameters to be estimated by Ordinary Least Squares (OLS) regression, such that the sum of the squared differences between observed values and the outcome predictions are minimised. Fixed effects are considered for the year (season) $\psi_{S(j)}$. Standard errors are clustered by tournament and match, where each tournament is considered a unique event, so for example, Wimbledon 2016 and Wimbledon 2017 are regarded as two separate tournaments. 

The three factors are calculated as follows: 

\vspace{2mm}
{\bf1. Odds implied probability:} 
The market's current implied probability of match outcome is derived from betting odds in the market. The implied probability $z_{ij}$ of player $i$ winning a match against player $j$ is calculated using inverse decimal odds:
\begin{equation}\label{eq:decimal-odds}
    z_{ij} = \frac{1}{odds_{ij}}
\end{equation}
where average bookmaker odds are always used for the market-implied probability (however, different odds are used for calculating returns when implementing the betting strategy that makes use of the mispricing model; we return to this in Section~\ref{sec:rep_estimates}). 
We follow RRS' implementation of using raw betting odds to derive probability $z_{ij}$. Therefore, for any match, $z_{ij} + z_{ji} = 1 + K > 1$, where K is the overround (i.e., the margin bookmakers gain from offering each side of an outcome). There are alternative methods that adjust implied odds for bookmaker overround \cite[e.g.,][]{shin-93,strumbelj-2014}.

\vspace{2mm}
{\bf 2. Rank Distance to Opponent:} 
The relative ability of opponents is measured using current WTA world rankings, which are derived from points obtained during tournament performances over the previous 52 weeks. 
\cite{BettingOnABuzz} justifiably suggest that ``the difference in ability between the 1st and 2nd ranked players is likely to be more than between the 100th and 101st ranked players''. To capture this effect, rank distance between opponents $i$ and $j$ is calculated as:
\begin{equation}\label{eq:rankdistrep}
    \textup{RankDist}_{ij} = -\left (  \frac{1}{{rank_{i}}} - \frac{1}{{rank_{j}}} \right )
\end{equation}
where unranked players are given an inverse rank of 0. 

\vspace{2mm}
{\bf 3. Wiki Relative Buzz Factor}:
A WikiBuzz factor is used to describe the recent increase in player interest from the crowd, measured by page views of player profiles on Wikipedia. The WikiBuzz factor for player $i$ against player $j$ is calculated as:
\begin{equation}
    \textup{WikiBuzz}_{ij}=\ln \left ( \frac{w_{i}}{\widetilde{w_{i}}} \right ) - \ln \left ( \frac{w_{j}}{\widetilde{w_{j}}} \right ) 
\end{equation}
where $w_{i}, w_{j}$ are the previous day's page views and  $\widetilde{w_{i}}, \widetilde{w_{j}}$ are the median daily page views over the past year, for players $i$ and $j$, respectively (see \ref{app:wikibuzz-example} for a worked example).

\begin{table}[t]
\renewcommand{\arraystretch}{1.1}
\centering
\caption{Replication of model estimates of betting market mispricing for WTA match results, 2015-2018: in-sample period only. RRS denotes original estimates of Equation (\ref{eq:errormodel}) \cite[Table~1, Col.~(IV)]{BettingOnABuzz}; CC denotes replication estimates.}
\label{tab:rep_estimates}
\tiny
\begin{tabular}{lSSSS}
\toprule
        & \multicolumn{2}{c}{\bf PM}      & \multicolumn{2}{c}{\bf PM w/o RD} \\ \cmidrule(lr){2-3}\cmidrule(l){4-5}
        & RRS  & \textbf{CC}  & RRS   & \textbf{CC}  \\ \midrule
Odds-implied probability, $\hat{\beta_{1}}$                & 0.025         & 0.025         & n/a         & 0.005       \\
                                        & (0.029) & (0.029) & n/a         & (0.026)     \\
WTA rank distance to opponent (RD), $\hat{\beta_{2}}$      & 0.055         & 0.055         &          &             \\
                                        & (0.029) & (0.029) &          &             \\
Wiki relative buzz factor, $\hat{\beta_{3}}$               & 0.009         & 0.009         & n/a         & 0.010       \\
                                        & (0.004)       & (0.004)       & n/a        & (0.004)     \\
Constant, $\hat{\alpha}$                                & -0.043  & -0.043   & n/a         & -0.033      \\
                                        & (0.015)       & (0.015)       & n/a         & (0.014)     \\ \midrule
\multicolumn{1}{l}{N of player-matches} & \multicolumn{1}{c}{~15,854}         & \multicolumn{1}{c}{~15,854}         & \multicolumn{1}{c}{~15,854}     & \multicolumn{1}{c}{~15,854}       \\ \bottomrule
\end{tabular}
\begin{tablenotes}
  \item Notes. Standard errors of estimates shown in parentheses.
  \item RRS: Ramirez-Reade-Singleton estimates \cite[copied from][Table~1, Column~(IV)]{BettingOnABuzz}.
  \item CC: \ifnum\BLIND=1{Estimates }\else{Clegg-Cartlidge estimates }\fi from replication study.
  \item PM: Error estimation according to ``Preferred model'' of \cite{BettingOnABuzz}, described by Equation~(\ref{eq:errormodel}).
  \item PM w/o RD: Error estimation according to Preferred model estimated without rank distance to opponent, 
  
  i.e., $\widetilde{y}_{ij} = \hat{\alpha} +\hat{\beta_{1}}z_{ij}+ \hat{\beta_{3}}\textup{WikiBuzz}_{ij}$.
  \item n/a: PM w/o RD model estimates are not reported in \cite{BettingOnABuzz}.
\end{tablenotes}
\end{table}

\vspace{2mm}
\noindent 
{\bf Results:} 
Table~\ref{tab:rep_estimates} presents replication results of the application of the \cite{mincer69eval} forecast evaluation framework.\footnote{We focus on two models, although we confirm that we are able to exactly reproduce all results presented in Table 1 of \cite{BettingOnABuzz}.} RRS indicates values are copied directly from the original paper \cite[Table 1, Column (IV)]{BettingOnABuzz} and CC indicates our replication results. Results for the preferred model (PM) are indentical, such that WikiBuzz and RankDist coefficients are significantly non-zero. We also provide our replication estimates for model PM w/o RD (the preferred model without rank distance), however, as these parameter estimates are not provided in the original paper, we cannot directly compare them against RRS' values. It can be seen that the WikiBuzz coefficient is also significantly non-zero for PM w/o RD. Therefore, our replication supports RRS' finding of the existence of bookmaker mispricing. 

\subsection{Market inefficiency}\label{sec:rep_estimates}
\noindent
RRS then test market inefficiency by simulating betting strategies that utilise the bookmaker mispricing model.

The outcome of a tennis match can be directly observed post hoc.
Following RRS, we let $y_{ij}=0$ if player $i$ loses to player $j$; and $y_{ij}=1$ if player $i$ wins against player $j$. Then, we can estimate the probability $\widetilde{y}_{ij}$ of player $i$ winning against player $j$ using a linear model with three factors: 

\begin{equation}\label{eq:pm}
    \widetilde{y}_{ij} = \hat{\alpha} +( 1+ \hat{\beta_{1}})~z_{ij}+ \hat{\beta_{2}}~\textup{RankDist}_{ij} + \hat{\beta_{3}}~\textup{WikiBuzz}_{ij}
\end{equation}

We refer to Equation~(\ref{eq:pm}) as the ``preferred model'' (PM). A second model, which does not include $\textup{RankDist}_{ij}$, is also explored. We refer to this alternative model as the ``preferred model without rank distance'' (PM w/o RD):
\begin{equation}\label{eq:pmword}
    \widetilde{y}_{ij} = \hat{\alpha} +( 1+ \hat{\beta_{1}})~z_{ij}+ \hat{\beta_{3}}~\textup{WikiBuzz}_{ij}
\end{equation}

Since the odds implied probability $z_{ij}$ represents the market's prediction of outcome probability plus some unknown overround, it is likely to be reasonably accurate and contain much greater predictive power than RankDist or WikiBuzz. Therefore, while the values of $\hat{\beta_{1}}$ are reported in the model estimates (see Table~\ref{tab:rep_estimates}), since the betting models of Equation (\ref{eq:pm}) and Equation (\ref{eq:pmword}) are predicting the outcome rather than estimating the conditional mean effects on bookmaker forecast errors, $(1+ \hat{\beta_{1}})$ is used as the multiplier for $z_{ij}$.

To implement a betting strategy, it is important to manage bet size and bankroll. The Kelly criterion, shown in Equation~(\ref{eq:kelly}), is a common strategy used to calculate $ f^{*}_{ij}$, the fraction of the current bankroll to bet on player $i$ to win against player $j$:

\begin{equation}\label{eq:kelly}
    f^{*}_{ij}=max\left( \widetilde{y}_{ij} -\frac{1-\widetilde{y}_{ij} }{b_{ij}} , 0\right)
\end{equation}

\noindent
where $\widetilde{y}_{ij} $ is the expected probability of $i$ to beat $j$  (estimated by the underlying betting model, e.g., PM given in Equation~(\ref{eq:pm}) or PM w/o RD given in Equation~(\ref{eq:pmword})) and $b_{ij}$ is the bet payoff (i.e., the bookmaker's decimal odds minus one).
If the expected win probability $\widetilde{y}_{ij}$ is significantly higher than the bookmaker's implied probability of a win, then the strategy will bet in an attempt to take advantage of a perceived bookmaker mispricing. Otherwise, the strategy will choose to not bet (i.e., $f^{*}_{ij}=0$). 

In sports betting research, the bankroll is typically reset to 1 before each bet, as documented in several studies (e.g., \cite{hvattum2010using}, \cite{boshnakov2017bivariate}, \cite{da2022forecasting}, \cite{Holmes_etal_2023}). We use this interpretation of the Kelly criterion in replicating the findings of \cite{BettingOnABuzz}.
Then, the profit (or loss) for the bet is calculated as: 
\begin{equation}
\label{eq:pnl}
  \pi_{ij}=\begin{cases}
    (f^{*}_{ij}\times odds) - f^{*}_{ij}, & \text{if $y_{ij}=1$}.\\
    -f^{*}_{ij}, & \text{if $y_{ij}=0$}.
  \end{cases}
\end{equation}

The primary metric used for evaluating a betting strategy's performance is the bettor's return on investment (ROI) over $N=2J$ potential bets, expressed as a percentage of the total amount bet over the sample period:
\begin{equation}
\label{eq:roi}
  \textup{ROI}=\frac{ \sum_{ij}^{2J}\pi_{ij} }{ \sum_{ij}^{2J}f^{*}_{ij} }
\end{equation}

An alternative metric is absolute returns over the sample period, where the final bankroll $B_T$ is expressed as a multiple of initial bankroll $B_0$, i.e.:
\begin{equation}
\label{eq:returns}
  \textup{Absolute returns}=\frac{ B_T }{ B_0 }
\end{equation}

Following RRS, we consider two alternative betting strategies to assess the performance of the models over the out-of-sample test period (January 2019 - February 2020, inclusive). The strategies use different betting odds for the calculation of the Kelly criterion (Equation~(\ref{eq:kelly})): the `Best Odds' strategy uses the best pre-match odds available in the market (as recorded by {\tt oddsportal.com}); while the `Bet365' strategy uses only pre-match odds available on Bet365. 

Results from the Bet365 strategy are shown in Table~\ref{tab:bet365-correction}. As before, RRS indicates values are copied directly from the original paper \cite[Table 4, Columns (II)-(IV)]{BettingOnABuzz} and CC indicates our replication results. It can be seen that we are able to replicate the original results when using the same dataset as RRS for training and testing, and we are able to confirm that the resultant series of bets placed are identical to RRS.\footnote{For completeness, in \ref{app:best-odds}, we also show an exact replication of results from betting strategies that use best odds in the market, even though ``using the best available odds just before a match begins is not normally realistic'' \citep{BettingOnABuzz}.}

\begin{figure}[t]
    \begin{subfigure}{.5\textwidth}
        \centering
        \includegraphics[width=.9\textwidth]{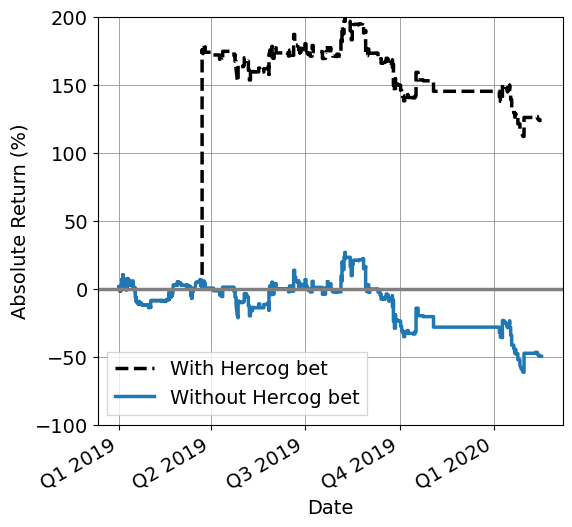}
        \caption{Preferred Model (PM)}
        \label{subfig:pm_with_hercog}
    \end{subfigure}%
        \begin{subfigure}{.5\textwidth}
        \centering
        \includegraphics[width=.9\textwidth]{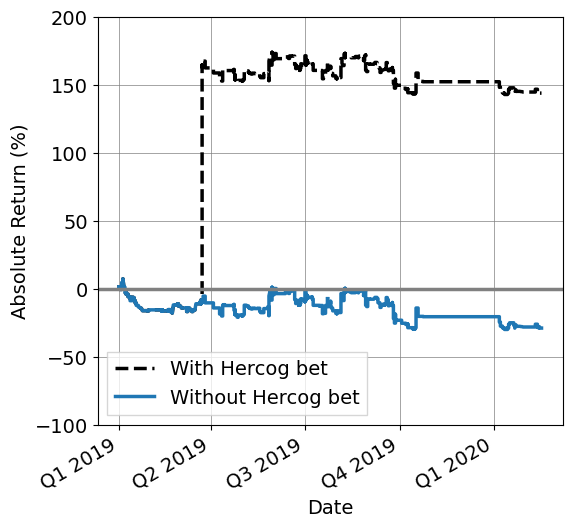}
        \caption{Preferred Model w/o RankDist (PM w/o RD)}
        \label{subfig:pmword_with_hercog}
    \end{subfigure}\\
\caption{Cumulative profits and the effect of the Hercog bet. When using Bet365 odds, the majority of reported out-of-sample returns \cite[Table 4]{BettingOnABuzz} are generated from a single bet on Hercog to win against Doi (March 22, 2019; see dashed line). When the Hercog bet is removed, both models make a loss (see solid line).}
\label{fig:hercogbet}
\end{figure}

However, for the best-performing betting models, we find that a large proportion of profits are generated by a single bet (the ``Hercog'' bet) on Bet365, which results from an inconsistency (i.e., an {\em error}) in the odds data provided by {\tt tennis-data.co.uk}.\footnote{RRS' dataset contains WTA data sourced from {\tt tennis-data.co.uk} for the period July 2015 to February 2020, inclusive; which includes information on every tennis match (competitors, competitor rankings, tournament, location, date, result, etc.) and a selection of pre-match betting odds taken from the major bookmakers, such as Bet365, as well as the market's average odds and best odds taken from the $K$ individual bookmakers listed on {\tt oddsportal.com} (where $K$ is normally in the range 40-60). A full description of the tennis data is available online: \href{http://www.tennis-data.co.uk/notes.txt}{tennis-data.co.uk/notes.txt}. For Hercog to beat Doi, {\tt oddsportal.com}  \href{https://www.oddsportal.com/tennis/usa/wta-miami-2019/hercog-polona-doi-misaki-8Wz8Rx2g/}{reports odds of 2.10} whereas {\tt tennis-data.co.uk} reports outlier odds of 5.50. Therefore, the Hercog error likely originates from {\tt tennis-data.co.uk}.}
In Figure~\ref{fig:hercogbet}, we plot the cumulative profit and loss graphs for two of the most profitable betting models proposed in the original paper \cite[Table 4, Column (III) and Column (IV)]{BettingOnABuzz}. The first, which we refer to as the ``preferred model'' (PM), has reported returns of $17.3\%$ when using Bet365 odds; and the second, which is a variation of the preferred model {\em without} the rank distance to opponent metric (PM w/o RD), has reported returns of $28.8\%$ on Bet365 odds.

The sudden surge in cumulative returns to over 150\% is due to a single large bet (of size 39.3\% bankroll) on Polona Hercog to win against Misaki Doi in the 2019 Miami Open. Despite the data recording the highest odds on Hercog to win (collected from a sample of between 40-60 bookmakers) as 2.10, and the market's average odds as 1.85, the Bet365 odds are listed as 5.50. This is a clear contradiction: the Bet365 odds on Hercog to win are an error.\footnote{Almost certainly, the Bet365 value is incorrect as it is an extreme outlier. The raw WTA data for 2019 (\href{http://www.tennis-data.co.uk/2019w/2019.xlsx}{http://www.tennis-data.co.uk/2019w/2019.xlsx}) contains 2,453 rows. For eventual winners, Bet365 win odds have minimum 1.01 and maximum 9; with mean 1.84 and standard deviation 0.80. Hercog's win odds of 5.5 are the 14th longest odds of the year, despite Hercog (rank 93) being favoured by the market to beat Doi (rank 112). All 13 rows with odds longer than 5 are for matches where rank outsiders beat a favourite.} Presented with these extremely long odds, the Kelly criterion anticipates significant bookmaker mispricing and specifies a proportionately large stake, which results in abnormal returns when Hercog wins.\footnote{In comparison, a unit betting strategy (i.e., place £1 on every mispriced bet) would exhibit a much less pronounced spike in returns from the Hercog error.} This affects the majority of the Bet365 betting results presented by RRS, and requires a correction.

\begin{table}[t]
\renewcommand{\arraystretch}{1.1}
\caption{Corrected out-of-sample Bet365 strategy results for WTA, 2019-2020.}
\label{tab:bet365-correction}
\vspace{1mm}
\tiny
\centering
\begin{tabular}{lS[table-format=4.2]S[table-format=4.2]S[table-format=4.2]lS[table-format=4.2]S[table-format=4.2]S[table-format=4.2]}
\toprule
                                   & \multicolumn{7}{c}{\bf Bet365 Odds}                           \\ \cmidrule(l){2-8}
                                   & \multicolumn{3}{c}{PM} & & \multicolumn{3}{c}{PM w/o RD} \\ \cmidrule(l){2-4}\cmidrule(l){5-8}
                                   & RRS     & CC    & CC/H   & & RRS    & CC    & CC/H    \\ \midrule
N odds ($2 \times J$ matches)      & 5156   & 5156   & 5155  & & 5156   & 5156   & 5155   \\
Number of bets placed              & 312     & 312    & 311  &  & 276    & 276    & 275    \\
Mean overround (\%)                & 6.46    & 6.46   & 6.46  & & 6.46   & 6.46   & 6.46   \\
Investment ($\times$ per bet budget)      & 7.15    & 7.15   & 6.77  & & 4.99   & 4.99   & 4.60   \\
Absolute return ($\times$ per bet budget) & 1.24    & 1.24  & -0.50 & & 1.44   & 1.44  & -0.29  \\
Return on Investment (\%)          & 17.26   & 17.29  & -7.36 & & 28.82  & 28.82  & -6.31  \\ \bottomrule
\end{tabular}
\begin{tablenotes}
  \item 1. Notes: RRS shows results copied from original paper: PM \cite[Table 4, Column III]{BettingOnABuzz}; PM w/o RD \cite[Table 4, Column IV]{BettingOnABuzz}. CC: Strict replication. CC/H: Replication without Hercog bet.
  \item 2. Replication: CC demonstrates an exact replication, apart from the ROI of PM. The original value 17.26 is a typo. 
  \item 3. Hercog bet: CC/H shows the impact of removing the single Hercog bet. ROI is negative for both models. 
\end{tablenotes}
\end{table}

\section{Correction}\label{sec:corrections}
\noindent
Table~\ref{tab:bet365-correction} shows corrected results of the most profitable models presented in RRS, Table~4. RRS shows the original published values; CC shows replication results on the original dataset (including the Hercog bet); and CC/H shows corrected results when the Hercog row is removed from the dataset. We see that strict replication results (CC) are the same as the original (RRS).\footnote{The only difference is ROI for PM. We have confirmed that the RRS reported value of $17.26$ is a simple typo, as RRS' original Stata code also indicates ROI value of $17.29$.} However, when the Hercog bet is removed (CC/H), the exceptional ROIs of $17.29\%$ and $28.82\%$ for PM and PM w/o RD, respectively, fall to negative values (i.e., {\em losses}) of $-7.36\%$ and $-6.31\%$. Consequently, final absolute returns are negative ($-50\%$ and $-29\%$), as shown in the cumulative profits plotted in Figure~\ref{fig:hercogbet}. 

\begin{table}[t]
\renewcommand{\arraystretch}{1.1}
\caption{Corrected out-of-sample Bet365 strategy results for WTA, 2019-2020: 
selecting sample average bookmaker odds, $p$, based on match competitiveness.}
\label{tab:table5corrected}
\vspace{1mm}
\tiny
\centering
\begin{tabular}{lS[table-format=4.2]S[table-format=4.2]S[table-format=4.2]lS[table-format=4.2]S[table-format=4.2]S[table-format=4.2]}
\toprule
                                   & \multicolumn{7}{c}{\bf Bet365 Odds}                           \\ \cmidrule(l){2-8}
                                   & \multicolumn{3}{c}{PM w/o RD: $p\in[0.2,0.8]$} & & \multicolumn{3}{c}{PM w/o RD: $p\in[0.4,0.6]$} \\ \cmidrule(l){2-4}\cmidrule(l){5-8}
                                   & RRS     & CC    & CC/H   & & RRS    & CC    & CC/H    \\ \midrule
N odds ($2 \times J$ matches)      & 4424   & 4454   & 4453  & & 1697   & 1697   & 1696   \\
Number of bets placed              & 363     & 384    & 383  &  & 263    & 263    & 262    \\
Mean overround (\%)                & 6.58    & 6.58   & 6.58  & & 7.01   & 7.01   & 7.01   \\
Investment ($\times$ per bet budget)      & 7.25    & 7.95   & 7.57  & & 9.27   & 9.27   & 8.91   \\
Absolute return ($\times$ per bet budget) & 1.46    & 1.44  & -0.25 & & 2.72   & 2.72  & 1.11  \\
Return on Investment (\%)          & 20.11   & 18.13  & -3.36 & & 29.38  & 29.38  & 12.44  \\ \bottomrule
\end{tabular}
\begin{tablenotes}
  \item 1. Notes: RRS shows results of PM w/o RD copied from original: $p\in[0.2,0.8]$ \cite[Table 5, Col. III]{BettingOnABuzz}; $p\in[0.4,0.6]$ \cite[Table 5, Col. IV]{BettingOnABuzz}. CC: Replication. CC/H: Replication without Hercog bet.
  \item 2. Replication: CC demonstrates a close replication. Minor differences in results for PM w/o RD are caused by a floating point error at the upper boundary $p=0.8$, which causes 30 potential bets to be overlooked in the original.  
  \item 3. Hercog bet: CC/H shows the impact of removing the single Hercog bet. Strategy $p\in[0.2,0.8]$ has negative ROI. However, for strategy $p\in[0.4,0.6]$, ROI falls but remains significantly positive.
\end{tablenotes}
\end{table}

Table~\ref{tab:table5corrected} shows corrected results of the models presented in RRS, Table~5. These betting strategies explore the effects of match competitiveness by only sampling matches (for model estimation and betting) with average odds implied probability $p$ within some interval range. When $p=0.5$, the market's average odds suggest an even match, with no favourite. As $p$ approaches $0$ or $1$, there is a clear favourite. For Hercog, $p=0.54$; i.e., this is a highly ``competitive'' match and Hercog is only a very slight favourite over Doi (despite the erroneous Bet365 odds of 5.5 listed on {\tt tennis-data.co.uk}). Therefore, RRS, Table~5, Columns III (PM w/o RD: $p\in[0.2,0.8]$) and IV (PM w/o RD: $p\in[0.4,0.6]$) are affected by the Hercog bet. In Table~\ref{tab:table5corrected}, CC shows replicated results, including the Hercog bet.\footnote{For completeness, in \ref{app:selectivestrategycols1and2}, we show the replicated results of the two columns unaffected by the Hercog bet, neither of which are significantly profitable.} Note that, while replication results (CC) are close to the original (RRS), there are some minor differences due to RRS's Stata code containing a floating point precision error for value $p=0.8$.\footnote{120 odds (90 in-sample; 30 out-of-sample) that should be $p=0.8$ are incorrectly stored by RRS as $0.800000011920929$. We include these odds for strategy $p\in[0.2,0.8]$.} As a result, our replication places an additional 21 bets and produces a slightly lower ROI than RRS. Much more significantly, when the Hercog bet is removed (CC/H), the strategy that samples $p\in[0.2,0.8]$ makes a loss. However, for $p\in[0.4,0.6]$, i.e., when only the most competitive matches are sampled for training and testing, ROI remains positive, albeit at a much lower rate of $12.44\%$ (previously, reported as $29.38\%)$. 

To assess the informative value of the 262 bets placed by the remaining profitable strategy, we refer to the betting return significance framework proposed by \cite{wunderlich2020betting} and calculate the probability $p_{bs}$ that a random betting strategy would have produced the same or a higher return as this positive ROI strategy. Here, we find $p_{bs}=0.002$; i.e., in 100,000 repeated simulation trials of randomly placing 262 bets in our dataset, 0.2\% of all random strategies generated equal or higher returns. Therefore, after Hercog error correction, we confirm that a profitable Bet365 strategy exists and it is unlikely to be a result of chance alone. While we note that the possibility of errors in the Wikipedia page view data could impact the reliability of this result, these consistent profits provide an indication of the existence of market inefficiencies.

\section{Exploration of market inefficiency and bookmaker mispricing over an extended dataset}\label{sec:extension}
\noindent
Here, we extend the original work of RRS by exploring the robustness of strategy PM w/o RD for competitive matches $p\in[0.4,0.6]$ on Bet365 odds over an extended period; and we explore bookmaker mispricing over this extended period using the preferred model (PM).

\subsection{New Dataset}\label{sec:new_data}
\noindent
Numerous tennis forecasting studies make use of data made available by {\tt tennis-data.co.uk} \cite[e.g.,][]{mchale2011bradley,candila2020neural,angelini2022weighted}. However, our replication of \cite{BettingOnABuzz} has identified that the raw data requires cleaning. 

Here, we generate an extended and cleaned data set, by using some basic filters to remove contradictory data entries, missing values, and outliers (see~\ref{app:data-processing} for details). We first drop rows where Bet365 odds are higher than the best odds available in the market. We then drop rows with any missing values. Finally, we drop rows with suspiciously high best odds, such that the absolute difference between the best odds in the market and average odds of the market is more than four times the standard deviation of the mean difference between best odds and average odds. 

The final dataset contains 35,274 rows (where each row contains odds for a player to win, also referred to as a player-match), covering 17,751 tennis matches and 584 players. Match dates range from 03 July 2015 to 31 August 2023, inclusive. We use matches between 03 July 2015 to 31 December 2018 for in-sample training; and 01 January 2019 to 31 August 2023 for out-of-sample testing.\footnote{
Cleaned data is available online: \href{https://github.com/Faxulous/notFeelingTheBuzz}{https://github.com/Faxulous/notFeelingTheBuzz}.}

\begin{figure}[t]
\centering
    \begin{subfigure}{.5\textwidth}
        \centering
        \includegraphics[width=.9\textwidth]{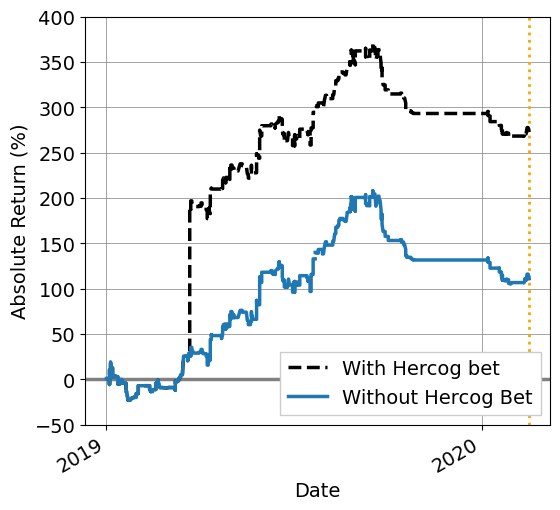}
        \caption{Dataset from \cite{BettingOnABuzz}.}
        \label{subfig:hercogselectivemodel}
    \end{subfigure}%
    \begin{subfigure}{.5\textwidth}
        \centering
        \includegraphics[width=.9\textwidth]{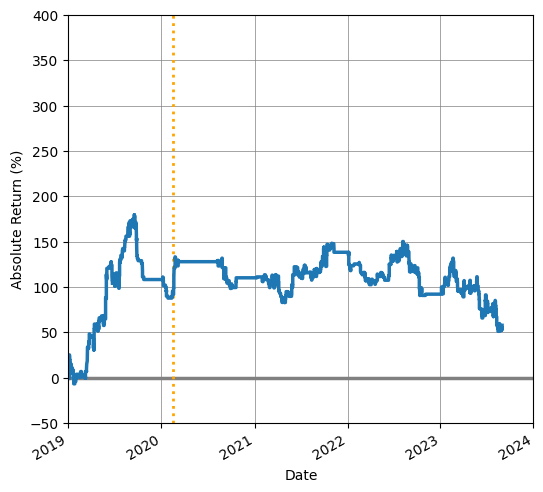}
        \caption{Cleaned and extended dataset.}
        \label{subfig:extendselectivemodel}
    \end{subfigure}%
\caption{Cumulative profits, using PM w/o RD: $p\in[0.4,0.6]$, shown in Table~\ref{tab:table5corrected}: (a) Original out-of-sample period (Jan. 2019 to Feb. 2020); (b) Extended out-of-sample period (Jan. 2019 - Aug. 2023). Dotted vertical line indicates change in dataset.}
\label{fig:selectivemodel}
\end{figure}

\subsection{Market inefficiency}
\noindent
We tested the strategy (using the original model coefficients) over an extended out-of-sample period. The cumulative profit and loss graph is shown in Figure~\ref{fig:selectivemodel}. On the left (Figure~\ref{subfig:hercogselectivemodel}), we see profits generated over the original out-of-sample period (Jan. 2019 -- Feb. 2020) with and without the Hercog bet. On the right (Figure~\ref{subfig:extendselectivemodel}), using the cleaned and extended dataset, we see that no further profits are generated after Feb. 2020. 
This potentially suggests that markets have become more efficient, although other unknown factors may
contribute to this apparent loss of profitability.

\subsection{Bookmaker mispricing} 
\noindent
Finally, we tested bookmaker mispricing by replicating the original mispricing test \cite[Table 1, Column IV]{BettingOnABuzz} on the preferred model (PM) over the period Feb. 2020 -- Aug. 2023. Shown in Table~\ref{tab:ext_estimates}, we find that new estimated coefficients for WikiBuzz and RankDist are no longer significant over this period, with respective p-values of $0.837$ and $0.471$. 

\begin{table}[t]
\renewcommand{\arraystretch}{1.1}
\centering
\caption{New model estimates of betting market mispricing for WTA match results, using extended out-of-sample period: Feb. 2020 -- Aug. 2023.}
\label{tab:ext_estimates}
\tiny
\begin{tabular}{lSSS}
\toprule
        & \multicolumn{2}{c}{\bf PM} \\ \cmidrule(l){2-3}
        & {Parameter} & {P-value} \\ \midrule
Odds-implied probability, $\hat{\beta_{1}}$                & 0.039 & 0.148        \\
                                        & (0.027) & \\
WTA rank distance to opponent (RD), $\hat{\beta_{2}}$      & 0.021 & 0.471       \\
                                        & (0.030) & \\
Wiki relative buzz factor, $\hat{\beta_{3}}$               & 0.001 & 0.837        \\
                                        & (0.004) &       \\
Constant, $\hat{\alpha}$                                & -0.046 & 0.001  \\
                                        & (0.014) &       \\ \midrule
\multicolumn{1}{l}{N of player-matches} & \multicolumn{2}{c}{~13,913}         \\ \bottomrule
\end{tabular}

\begin{tablenotes}
  \item Notes. Standard errors of parameter estimates are shown in parentheses.
  \item PM: Error estimation according to ``Preferred model'' of \cite{BettingOnABuzz}, described by Equation~(\ref{eq:errormodel}).
\end{tablenotes}
\end{table}

We present this as tentative evidence that, since 2020, WikiBuzz and RankDist factors are no longer reliable predictors of bookmaker mispricing.

\section{Discussion}\label{sec:discussion} 
\noindent
While \cite{BettingOnABuzz} consider Wikipedia as a predictor for crowd wisdom in online data, other sources of crowd wisdom for sports forecasting have been investigated elsewhere. For example, \cite{luckner-08-prediction-markets} and \cite{spann-2009-prediction-markets} find that prediction markets have improved forecast accuracy for FIFA World Cup and Bundesliga soccer outcomes, respectively; \cite{schumaker2016predicting} and \cite{brown-2018-social-media} find the aggregate sentiment/tone of Tweets can improve the forecast accuracy of English Premier League soccer matches; \cite{peeters2018testing} finds that crowd valuations of professional soccer players on {\tt tranfermarkt.de} improves forecast accuracy of international soccer matches; and \cite{brown-2019-amateur-crowds} find aggregated tips from amateur tipsters on {\tt oddsportal.com} improves forecast accuracy across a variety of sports. 

Also, like \cite{BettingOnABuzz}, many of these studies evaluate their sports forecasting models using historical bookmaker odds to simulate betting returns \cite[e.g.,][]{spann-2009-prediction-markets,schumaker2016predicting,brown-2018-social-media,peeters2018testing,brown-2019-amateur-crowds}. However, this raises three issues: (i) as average bookmaker odds are known to be a powerful predictor in their own right \cite[e.g.,][]{vstrumbelj2010online}, it is possible for betting profits to be generated by simply exploiting bookmaker odds that are significantly larger than the mean; (ii) as demonstrated by \cite{wunderlich2020betting}, betting returns are not a valid measure of forecasting model accuracy; (iii) as we have shown in Section~\ref{sec:replication}, betting returns are particularly sensitive to outlier odds and even well-regarded public repositories such as {\tt tennis-data.co.uk} are not impervious to data errors. 

Therefore, in light of our findings, we suggest that many works in the area of sports forecasting -- particularly those that present betting returns -- should be prioritised for replication; while related third-party data sources used for these studies should be quality checked and cleaned.

\section{Conclusions}\label{sec:conclusion}
\noindent
We have replicated and corrected an error in the work of \cite{BettingOnABuzz}. 
While the correction significantly affects reported betting returns,
when considering the original data period our replication findings do not contradict the original authors' claims of bookmaker mispricing and market inefficiencies. In particular, our finding of significant profitability for the Bet365 selective strategy based on ``competitive'' matches (Section~\ref{sec:corrections}) may indicate the existence of market inefficiencies. However, the introduction of a new data period (Section~\ref{sec:extension}) resulted in the loss of the model's predictive power for both inefficiency and mispricing, thus highlighting the need for further research to account for potential limitations of the model.

Our work exemplifies the sensitivity of betting models to data quality, 
and expands upon the findings of \cite{wunderlich2020betting}
to demonstrate that even a \emph{single} data error can significantly influence ROI when adopting a Kelly staking strategy. Therefore, simulated betting returns should be interpreted cautiously, and data quality should always be checked, even when using popular open source datasets such as {\tt tennis-data.co.uk}.

More generally, our findings also demonstrate the importance of replication studies in forecasting research. In this instance, having access to the original data has proven invaluable and we thank the authors for making their data and code available. The {\em International Journal of Forecasting} (IJF) -- the venue where the original work was published -- promotes replication \citep{Hyndman-2010-editorial-replication} and from July 2023 has made data and code sharing mandatory. IJF also publishes replication studies \citep[e.g.,][]{adya-2000-correction-paper,Boylan-15-reproducability}. 

However, as described at length by \cite{Makridakis-18-reproducibility}, not all journals take such an objective standpoint on replication and reproducibility. 
We, the authors, strongly agree with the imperative of replication 
and to help address this issue, one of us \ifnum\BLIND=1{(name redacted) }\else{(Cartlidge) }\fi encourages the use of replication studies when supervising undergraduate and postgraduate student projects. To this end, the preliminary research that formed the basis of this present paper was initially performed by \ifnum\BLIND=1{(name redacted) }\else{Clegg }\fi as part of his MSc project\ifnum\BLIND=1{. }\else{ in Financial Technology with Data Science at the University of Bristol. }\fi 
\ifnum\BLIND=1{In previous years, project students have had success publishing replication studies that reproduce and extend previous work (citation redacted), or provide some corrections to previous work (citations redacted). }\else{In previous years, students supervised by Cartlidge have had success publishing replication studies that reproduce and extend previous work \citep[e.g.,][]{duffin-18-latency}, or provide some corrections to previous work \citep[e.g.,][]{clamp-13-pricingcloud,Cartlidge-14-correcting-brokerage,stotter-13-ASAD,stotter-14-expo}. }\fi Harnessing student research hours for replication studies is beneficial for all parties, but encouraging students to perform replication is challenging if their results cannot be published. Therefore, we commend the IJF's approach to reproducible research and we urge other journal venues to follow suit, if they do not already. 

\ifnum\BLIND=1{}
\else{
\section*{Acknowledgements}
\noindent This work was supported by UKRI grant \href{https://gow.epsrc.ukri.org/NGBOViewGrant.aspx?GrantRef=EP/Y028392/1}{EP/Y028392/1}: ``AI for Collective Intelligence (AI4CI) Research Hub''.
}\fi

\pagebreak
\bibliographystyle{elsarticle-harv} 
\ifnum\BBL=1
    \include{bbl.tex} 
\else
    \bibliography{cas-refs}
\fi

\pagebreak
\appendix

\section{WikiBuzz Example}\label{app:wikibuzz-example}

\begin{table}[ht]
\caption{Miami Open Pageviews}
\label{tab:wikibuzz-example}
\centering
\tiny
\begin{tabular}{lrlrl} 
\toprule
\multicolumn{1}{c}{} & \multicolumn{2}{c}{\textbf{Danielle Collins}}           & \multicolumn{2}{c}{\textbf{Jelena Ostapenko}}            \\ 
{} & \multicolumn{1}{c}{\bf Pageviews} & \multicolumn{1}{c}{\bf Match} & {\bf Pageviews} & \multicolumn{1}{c}{\bf Match}  \\ 
\cmidrule(r){2-3}\cmidrule(l){4-5}
15-Mar  & 212   &   & 27    &   \\ 
16-Mar  & 111   &   & 57    &   \\ 
17-Mar  & 86    &   & 55    &   \\ 
18-Mar  & 66    &   & 38    &   \\ 
19-Mar  & 188  & Qualifier   & 28  &   \\ 
20-Mar  & 246  & Qualifier  & 43   &  \\ 
21-Mar  & 565  & Round of 128   & 23    &   \\ 
22-Mar  & 380  &                & 19   &  \\ 
23-Mar  & 1,023    & Round of 64   & 35  & Round of 64     \\ 
24-Mar  & 827   &      & 20  &    \\ 
25-Mar  & 2,097   & Round of 32    & 39   & Round of 32     \\ 
26-Mar  & 2,485   &    & 36      &     \\ 
27-Mar  & 7,779   & Round of 16    & 44   & Round of 16     \\ 
28-Mar  & 12,208    &   & 54   & Quarter-final   \\ 
29-Mar  & 39,955    & Quarter-final  & 39    &    \\ 
30-Mar  & 21,777     & Semi-final     & 180    & Semi-final   \\
\hline
\end{tabular}
\end{table}

\noindent
Here, we present a worked example of Wikipedia Relative Buzz Factor for a semi-final match at the Miami Open between Danielle Collins and Jalena Ostapenko, on 30th March 2018.

Table~\ref{tab:wikibuzz-example} shows that the pageviews of Danielle Collins increase dramatically as she progresses through the tournament. In comparison, Jelena Ostapenko's pageviews never exceed 180. As an unranked and relatively obscure player, each of Danielle Collins' wins propelled more people to view her Wikipedia page, with a buzz developing as she unexpectedly survived ever deeper into the tournament. For the semi-final match, Collins has a relative WikiBuzz factor of $6.93$, calculated using her previous day's views of $39,955$ and median views over the previous year of $26.5$, against Ostapenko's previous day's views of $39$ and yearly median of $27$: 

\begin{equation}
\label{eq:wikkibuzz-example}
    6.93=\ln \left ( \frac{39955}{26.5} \right ) - \ln \left ( \frac{39}{27} \right ) 
\end{equation}

\vspace{2mm}
In the event, there was no fairy-tale ending for Collins, as she lost to Ostapenko in two sets.

\pagebreak
\section{Best Odds Strategy Results from Mispricing Test}
\label{app:best-odds}

\begin{table}[ht]
\renewcommand{\arraystretch}{1.1}
\caption{Corrected out-of-sample Best Odds strategy results for WTA, 2019-2020.}
\label{tab:table4bestodds}
\vspace{1mm}
\tiny
\centering
\begin{tabular}{lS[table-format=4.2]S[table-format=4.2]}
\toprule
                                   & \multicolumn{2}{c}{\bf Best Odds}   \\ \cmidrule(l){2-3}
                                   & \multicolumn{2}{c}{PM}               \\ \cmidrule(l){2-3}
                                   & {RRS}     & {CC}       \\ \midrule
N odds ($2 \times J$ matches)      & 5188   & 5189   \\
Number of bets placed              & 2350     & 2350    \\
Mean overround (\%)                & -0.23    & -0.24   \\
Investment ($\times$ per bet budget)      & 76.63    & 76.63   \\
Absolute return ($\times$ per bet budget) & 2.34    & 2.34  \\
Return on Investment (\%)          & 3.05   & 3.05  \\ \bottomrule
\end{tabular}

\begin{tablenotes}
  \item 1. Notes: RRS shows results copied from the original paper: PM \cite[Table 4, Column II]{BettingOnABuzz}. CC: Strict replication.
  \item 2. Replication: CC demonstrates an exact replication, apart from the N odds. The difference in $N$ of $1$ is due to a calculation error by RRS. 
\end{tablenotes}
\end{table}
\noindent
Betting results using best odds are almost exactly replicated, apart from a minor discrepancy of $N=5,188$ odds \cite[Table 4, Column (II)]{BettingOnABuzz} as opposed to the $N=5,189$ odds used in our replication. We traced the discrepancy in $N$ odds of $1$ and mean overround of $0.01$ to a calculation by RRS regarding the match between Anastasia Potapova and Kristina Kucova on July 15, 2019.

The best odds for Kucova are unavailable, but they are available for Potapova. When calculating the $N$ odds used, the logic used by \cite{BettingOnABuzz} doesn't count either bet. However, to obtain the results shown in Table~\ref{tab:table4bestodds} the Potapova bet \emph{must} be included, as it results in a difference in ROI of $0.02$. Therefore, the true value of $N=5189$ and the mean overround is $-0.24\%$.

The Hercog error corrected in Section~\ref{sec:corrections} is not present here, as the best odds (for the Polona Hercog vs. Misaki Doi match on March 22, 2019) are correctly listed as $2.09$ for Hercog to win.

\pagebreak
\section{Bet365 Selective betting strategy based on match competitiveness}
\label{app:selectivestrategycols1and2}

\begin{table}[ht]
\renewcommand{\arraystretch}{1.1}
\caption{Corrected out-of-sample Bet365 strategy results for WTA, 2019-2020: 
selecting sample average bookmaker odds, $p$, based on match competitiveness; additional columns.}
\label{tab:table5corrected-extracolumns}
\vspace{1mm}
\tiny
\centering
\begin{tabular}{lS[table-format=4.2]S[table-format=4.2]S[table-format=4.2]S[table-format=4.2]S[table-format=4.2]}
\toprule
                                   & \multicolumn{4}{c}{\bf Bet365 Odds}                           \\ \cmidrule(l){2-5}
                                   & \multicolumn{2}{c}{PM w/o RD: $p\in (0, 0.2)\cup(0.8, 1)$} & \multicolumn{2}{c}{PM w/o RD: $p\in(0, 0.4)\cup(0.6, 1)$} \\ \cmidrule(lr){2-3}\cmidrule(lr){4-5}
                                   & {RRS}    & {CC}    & {RRS}    & {CC}  \\ \midrule
N odds ($2 \times J$ matches)      & 732   & 702   & 3459   & 3459  \\
Number of bets placed              & 4    & 7    & 87   & 87  \\
Mean overround (\%)                & 5.71   & 5.70   & 6.02   & 6.18   \\
Investment ($\times$ per bet budget)      & 0.05   & 0.18   & 1.03   & 1.03   \\
Absolute return ($\times$ per bet budget) & -0.002  & -0.06  & 0.01   & 0.01  \\
Return on Investment (\%)          & -3.02  & -33.8  & 0.81   & 0.81  \\ \bottomrule
\end{tabular}
\begin{tablenotes}
  \item 1. Notes: RRS shows results, copied from original, using PM w/o RD selecting sample odds based on match competitiveness: $p\in (0, 0.2)\cup(0.8, 1)$ and $p\in(0, 0.4)\cup(0.6, 1)$ \cite[Table 5, Cols I, II]{BettingOnABuzz}. CC: Replication.
  \item 2. Replication: CC demonstrates a close replication. Minor differences in results for PM w/o RD are caused by a floating point error at the upper boundary $p=0.8$, which causes 30 potential bets to be overlooked in the original.  
  \item 3. We find the mean overround for $p\in(0, 0.4)\cup(0.6, 1)$ to be 6.18, misreported as 6.02 by RRS.
\end{tablenotes}
\end{table}
\noindent
Table~\ref{tab:table5corrected-extracolumns} shows the replication results of the two strategies from RRS, Table~5, that do not bet on the Hercog bet. The strategy PM w/o RD: $p\in(0, 0.4)\cup(0.6, 1)$ is marginally profitable with an ROI of $0.81\%$. However, with only 87 bets placed, we do not consider this profit to be significant. Using the framework proposed by \cite{wunderlich2020betting}, we can calculate the probability $p_{bs}$ that a random betting strategy would have produced the same or a higher return. Here, we find $p_{bs}=0.22$; i.e., in 100,000 repeated simulation trials of randomly placing 87 bets in our dataset, 22\% of all random strategies generated equal or higher returns. 

\pagebreak
\section{Data Cleaning Process for Extended Dataset}
\label{app:data-processing}

\begin{table}[ht]
\renewcommand{\arraystretch}{1.1}
\caption{Data cleaning steps for the extended dataset used in Section~\ref{sec:extension}. Each step shows an example row that is removed, including match date, player names $i$ and $j$, betting odds for $i$ to win, and earliest available Wikipedia profile data for each player. `Rows' indicates data size after processing step.}
\label{tab:data-cleaning}
\centering
\tiny
\begin{tabular}{cccllccccc} 
\toprule
& & \multicolumn{8}{c}{\textbf{Example Row Removed During Processing Step}} \\
\cmidrule{3-10}
& Rows & Date & \multicolumn{1}{c}{Player $i$} & \multicolumn{1}{c}{Player $j$} & Best$_i$ & Av.$_i$ & Bet365$_i$ & Wiki$_i$ & Wiki$_j$ \\
\midrule
0 & 37,700 \\
1 & 35,680 & 07/06/21 & E. Raducanu & H. Dart & 2.60 & 2.41 & 2.37 & \underline{05/06/21} & 01/07/15\\
2 & 35,648 & 23/03/19 & P. Hercog & M. Doi & 2.10 & 1.85 & \underline{5.50} & 01/07/15 & 01/07/15\\
3 & 35,474 & 06/04/21 & L. Arruabarrena  & J. Plazas & 1.04 & 1.01 & \underline{None} & 01/07/15 & 12/04/18\\
4 & 35,274 & 10/01/17 & C. Wozniacki & Y. Putintseva & \underline{127} & 1.24 & 1.22 & 01/07/15 & 01/07/15\\
\bottomrule
\end{tabular}
\end{table}

\noindent
The raw dataset contains data from 01 July 2015 to 31 August 2023. There are a total of 37,700 rows (N odds, which we refer to as player-matches), covering 18,850 tennis matches and 688 players. 

We process the raw data using the following steps (refer to Table~\ref{tab:data-cleaning}):
\begin{description}
    \item [1. New players:] To account for emerging players with no profile history, consider only matches where both players have a Wiki profile page that has existed for at least one year and one day. Rows removed: 2,020.
    \item [2. Bet365:] To account for inconsistencies in odds across bookmakers, remove rows where Bet365 odds are higher than best odds across the market, as reported by {\tt oddsportal}. Rows removed: 32. 
    \item [3. Missing odds:] Remove rows with missing odds. Rows removed: 174.
    \item [4. Best odds:] To account for abnormally long odds that have questionable integrity, remove rows where the difference between the implied probabilities of best odds and average odds is greater than four standard deviations of the mean difference between best odds and average odds. The standard deviation is calculated as 0.014. Rows removed: 200.
\end{description}
The final dataset contains 35,274 rows (N odds of player-matches), covering 17,751 tennis matches and 584 players. Match dates range from 03 July 2015 to 31 August 2023. Cleaned data is available online.\footnote{Cleaned data: \href{https://github.com/Faxulous/notFeelingTheBuzz}{https://github.com/Faxulous/notFeelingTheBuzz}.}

\pagebreak

\end{document}